\documentclass[twocolumn,preprint]{aastex631}

\usepackage{amsmath,amstext,natbib,apjfonts,longtable,verbatim,float,graphicx,color, longtable}

\newcommand{\target}{CEERS-AGN-z5-1} 
\defcitealias{onoue23}{O23}
\defcitealias{greene_ho05}{GH05}

\newcommand{\msun}{M_\odot}

\newcommand{\zsun}{Z_\odot}

\newcommand{\CII}{\hbox{{\rm [}{\rm C}~{\sc ii}{\rm ]}}}
\defcitealias{VB_2001}{VB01}

\newcommand{\HII}{\hbox{{\rm H}~{\sc ii}}}

\newcommand{\HeII}{\hbox{{\rm He}~{\sc ii}}}

\newcommand{\NII}{\hbox{{\rm [N}~{\sc ii}{\rm ]}}}
\newcommand{\OII}{\hbox{{\rm [O}~{\sc ii}{\rm ]}}}
\newcommand{\OIII}{\hbox{{\rm [O}~{\sc iii}{\rm ]}}}
\newcommand{\NeIII}{\hbox{{\rm [Ne}~{\sc iii}{\rm ]}}}

\newcommand{\NeV}{\hbox{{\rm [Ne}~{\sc v}{\rm ]}}}
\newcommand{\FeX}{\hbox{{\rm [Fe}~{\sc x}{\rm ]}}}

\newcommand{\Ha}{\hbox{{\rm H}$\alpha$}}
\newcommand{\Hb}{\hbox{{\rm H}$\beta$}}
\newcommand{\Hg}{\hbox{{\rm H}$\gamma$}}

\shorttitle{CEERS: Low-Mass, Broad-Line AGN at z>5}
\shortauthors{Kocevski et al.}
\graphicspath{{./}{figures/}}

\begin{document}

\title{\large \bf Hidden Little Monsters: Spectroscopic Identification of Low-Mass, Broad-Line AGN at $z>5$ with CEERS}

\suppressAffiliations

\author[0000-0002-8360-3880]{Dale D. Kocevski}
\affiliation{Department of Physics and Astronomy, Colby College, Waterville, ME 04901, USA}

\author[0000-0003-2984-6803]{Masafusa Onoue}
\altaffiliation{Kavli Astrophysics Fellow}
\affiliation{Kavli Institute for Astronomy and Astrophysics, Peking University, Beijing 100871, China}
\affiliation{Kavli Institute for the Physics and Mathematics of the Universe (Kavli IPMU, WPI), The University of Tokyo, Chiba 277-8583, Japan}

\author[0000-0001-9840-4959]{Kohei Inayoshi}
\affiliation{Kavli Institute for Astronomy and Astrophysics, Peking University, Beijing 100871, China}

\author[0000-0002-1410-0470]{Jonathan R. Trump}
\affiliation{Department of Physics, 196 Auditorium Road, Unit 3046, University of Connecticut, Storrs, CT 06269}

\author[0000-0002-7959-8783]{Pablo Arrabal Haro}
\affiliation{NSF's National Optical-Infrared Astronomy Research Laboratory, 950 N. Cherry Ave., Tucson, AZ 85719, USA}

\author[0000-0002-5688-0663]{Andrea Grazian}
\affil{INAF--Osservatorio Astronomico di Padova, 
Vicolo dell'Osservatorio 5, I-35122, Padova, Italy\\}

\author[0000-0001-5414-5131]{Mark Dickinson}
\affiliation{NSF's National Optical-Infrared Astronomy Research Laboratory, 950 N. Cherry Ave., Tucson, AZ 85719, USA}

\author[0000-0001-8519-1130]{Steven L. Finkelstein}
\affiliation{Department of Astronomy, The University of Texas at Austin, 2515 Speedway, Stop C1400, Austin, TX 78712, USA}

\author[0000-0001-9187-3605]{Jeyhan S. Kartaltepe}
\affiliation{Laboratory for Multiwavelength Astrophysics, School of Physics and Astronomy, Rochester Institute of Technology, 84 Lomb Memorial Drive, Rochester, NY 14623, USA}

\author[0000-0002-3301-3321]{Michaela Hirschmann}
\affiliation{Institute of Physics, Laboratory of Galaxy Evolution, Ecole Polytechnique Fédérale de Lausanne (EPFL), Observatoire de Sauverny, 1290 Versoix, Switzerland}

\author[0000-0001-7201-5066]{Seiji Fujimoto}
\affiliation{Department of Astronomy, The University of Texas at Austin, Austin, TX 78712, USA}
\affiliation{Cosmic Dawn Center (DAWN), Jagtvej 128, DK2200 Copenhagen N, Denmark}
\affiliation{Niels Bohr Institute, University of Copenhagen, Lyngbyvej 2, DK2100 Copenhagen \O, Denmark}

\author[0000-0002-0000-2394]{St{\'e}phanie Juneau}
\affiliation{NSF's NOIRLab, 950 N. Cherry Ave., Tucson, AZ 85719, USA}


\author[0000-0001-5758-1000]{Ricardo O. Amor\'{i}n}
\affiliation{Instituto de Investigaci\'{o}n Multidisciplinar en Ciencia y Tecnolog\'{i}a, Universidad de La Serena, Raul Bitr\'{a}n 1305, La Serena 2204000, Chile}
\affiliation{Departamento de Astronom\'{i}a, Universidad de La Serena, Av. Juan Cisternas 1200 Norte, La Serena 1720236, Chile}

\author[0000-0002-9921-9218]{Micaela B. Bagley}
\affiliation{Department of Astronomy, The University of Texas at Austin, Austin, TX, USA}
   
\author[0000-0002-0786-7307]{Guillermo Barro}
\affiliation{Department of Physics, University of the Pacific, Stockton, CA 90340 USA}

\author[0000-0002-5564-9873]{Eric F.\ Bell}
\affiliation{Department of Astronomy, University of Michigan, 1085 S. University Ave, Ann Arbor, MI 48109-1107, USA}

\author[0000-0003-0492-4924]{Laura Bisigello}
\affiliation{Dipartimento di Fisica e Astronomia "G.Galilei", Universit\'a di Padova, Via Marzolo 8, I-35131 Padova, Italy}
\affiliation{INAF--Osservatorio Astronomico di Padova, Vicolo dell'Osservatorio 5, I-35122, Padova, Italy}

\author[0000-0003-2536-1614]{Antonello Calabr{\`o}}
\affiliation{Osservatorio Astronomico di Roma, via Frascati 33, Monte Porzio Catone, Italy}

\author[0000-0001-7151-009X]{Nikko J. Cleri}
\affiliation{Department of Physics and Astronomy, Texas A\&M University, College Station, TX, 77843-4242 USA}
\affiliation{George P.\ and Cynthia Woods Mitchell Institute for Fundamental Physics and Astronomy, Texas A\&M University, College Station, TX, 77843-4242 USA}

\author[0000-0003-1371-6019]{M. C. Cooper}
\affiliation{Department of Physics \& Astronomy, University of California, Irvine, 4129 Reines Hall, Irvine, CA 92697, USA}

\author[0000-0001-8917-2148]{Xuheng Ding}
\affiliation{Kavli Institute for the Physics and Mathematics of the Universe (Kavli IPMU, WPI), The University of Tokyo, Chiba 277-8583, Japan}

\author[0000-0001-9440-8872]{Norman A. Grogin}
\affiliation{Space Telescope Science Institute, Baltimore, MD, USA}

\author[0000-0001-6947-5846]{Luis C. Ho}
\affiliation{Kavli Institute for Astronomy and Astrophysics, Peking University, Beijing 100871, China}
\affiliation{Department of Astronomy, School of Physics, Peking University, Beijing 100871, China}


\author[0000-0002-7779-8677]{Akio K. Inoue}
\affiliation{Waseda Research Institute for Science and Engineering, Faculty of Science and Engineering, Waseda University, 3-4-1, Okubo, Shinjuku, Tokyo 169-8555, Japan}
\affiliation{Department of Physics, School of Advanced Science and Engineering, Faculty of Science and Engineering, Waseda University, 3-4-1, Okubo, Shinjuku, Tokyo
169-8555, Japan}

\author[0000-0003-4176-6486]{Linhua Jiang}
\affiliation{Kavli Institute for Astronomy and Astrophysics, Peking University, Beijing 100871, China}
\affiliation{Department of Astronomy, School of Physics, Peking University, Beijing 100871, China}

\author{Brenda Jones}
\affiliation{Department of Physics and Astronomy, University of Maine, Orono, ME 04469-5709}

\author[0000-0002-6610-2048]{Anton M. Koekemoer}
\affiliation{Space Telescope Science Institute, 3700 San Martin Dr., Baltimore, MD 21218, USA}

\author[0000-0002-1044-4081]{Wenxiu Li}
\affiliation{Kavli Institute for Astronomy and Astrophysics, Peking University, Beijing 100871, China}

\author[0000-0002-8502-7573]{Zhengrong Li}
\affiliation{Kavli Institute for Astronomy and Astrophysics, Peking University, Beijing 100871, China}

\author[0000-0001-8688-2443]{Elizabeth J.\ McGrath}
\affiliation{Department of Physics and Astronomy, Colby College, Waterville, ME 04901, USA}

\author[0000-0002-8136-8127]{Juan Molina}
\affiliation{Department of Space, Earth and Environment, Chalmers University of Technology, 
Onsala Space Observatory, 439 92 Onsala, Sweden}
\affiliation{Kavli Institute for Astronomy and Astrophysics, Peking University, Beijing 100871, China}

\author[0000-0001-7503-8482]{Casey Papovich}
\affiliation{Department of Physics and Astronomy, Texas A\&M University, College Station, TX, 77843-4242 USA}
\affiliation{George P.\ and Cynthia Woods Mitchell Institute for Fundamental Physics and Astronomy, Texas A\&M University, College Station, TX, 77843-4242 USA}

\author[0000-0003-4528-5639]{Pablo G. P\'erez-Gonz\'alez}
\affiliation{Centro de Astrobiolog\'{\i}a (CAB), CSIC-INTA, Ctra. de Ajalvir km 4, Torrej\'on de Ardoz, E-28850, Madrid, Spain}

\author[0000-0003-3382-5941]{Nor Pirzkal}
\affiliation{ESA/AURA Space Telescope Science Institute}

\author[0000-0003-3903-6935]{Stephen M.~Wilkins} %
\affiliation{Astronomy Centre, University of Sussex, Falmer, Brighton BN1 9QH, UK}
\affiliation{Institute of Space Sciences and Astronomy, University of Malta, Msida MSD 2080, Malta}

\author[0000-0001-8835-7722]{Guang Yang}
\affiliation{Kapteyn Astronomical Institute, University of Groningen, P.O. Box 800, 9700 AV Groningen, The Netherlands}
\affiliation{SRON Netherlands Institute for Space Research, Postbus 800, 9700 AV Groningen, The Netherlands}

\author[0000-0003-3466-035X]{L. Y. Aaron\ Yung}
\affiliation{Astrophysics Science Division, NASA Goddard Space Flight Center, 8800 Greenbelt Rd, Greenbelt, MD 20771, USA}

\begin{abstract}
We report on the discovery of two low-luminosity, broad-line active galactic nuclei (AGN) at $z>5$ identified using JWST NIRSpec spectroscopy from the Cosmic Evolution Early Release Science (CEERS) Survey.   We detect broad H$\alpha$
emission in the spectra of both sources, with FWHM of $2038\pm286$ and $1807\pm207$ km s$^{-1}$, resulting in virial black hole (BH) masses that are 1-2 dex below that of existing samples of luminous quasars at $z>5$. The first source, CEERS 1670 at $z=5.242$, is 2--3 dex fainter than known quasars at similar redshifts and was previously identified as a candidate low-luminosity AGN based on its morphology and rest-frame optical spectral energy distribution (SED).  We measure a BH mass of $M_{\rm BH}=1.3\pm0.4\times 10^{7}~\msun$, confirming that this AGN is powered by the least-massive BH known in the universe at the end of cosmic reionization.  The second source, CEERS 3210 at $z=5.624$, is inferred to be a heavily obscured, broad-line AGN caught in a transition phase between a dust-obscured starburst and an unobscured quasar.  We estimate its BH mass to be in the range of $M_{\rm BH}\simeq 0.9-4.7 \times 10^{7}~M_{\odot}$, depending on the level of dust obscuration assumed.  We perform SED fitting to derive host stellar masses, $M_\star$, allowing us to place constraints on the BH-galaxy mass relationship in the lowest mass range yet probed in the early universe.  The $M_{\rm BH}/M_\star$ ratio for CEERS 1670, in particular, is consistent with or higher than the empirical relationship seen in massive galaxies at $z=0$.  We examine the narrow emission-line ratios of both sources and find that their location on the BPT and OHNO diagrams is consistent with model predictions for moderately low-metallicity AGN with $Z/Z_\odot \simeq 0.2-0.4$.  The spectroscopic identification of low-luminosity, broad-line AGN at $z>5$ with $M_{\rm BH}\simeq 10^{7}~\msun$ demonstrates the capability of JWST to push BH masses closer to the range predicted for the BH seed population and provides a unique opportunity to study the early stages of BH-galaxy assembly.

\end{abstract}


\section{Introduction}

With the advent of wide-field quasar surveys such as those carried out by the Sloan Digital Sky Survey \citep[SDSS;][]{fan01, jiang16} and the Panoramic Survey Telescope \& Rapid Response System 1 (Pan-STARRS1; \citealt{Banados_2016,Mazzucchelli_2017}), hundreds of quasars have been discovered and characterized at $z>5$ \citep{inayoshi20, fan22}, with the most distant found a mere 670 million years after the Big Bang \citep{Wang21}.  The super massive black holes (SMBHs) that power these sources have masses of order $\sim 10^{9}~\msun$, raising the question of how such systems were built in such a short amount of cosmic time. Most theories involve Eddington-limited or possibly super-Eddington accretion onto seed BHs that are  predicted to form at $10<z<30$ and have masses that range from $\sim 10^{2}~\msun$ (so called “light seeds”) to over $\sim10^{5} M_{\odot}$ (“heavy seeds”) with a continuous distribution \citep[e.g.,][]{inayoshi20, Volonteri_2021}. The relative contribution of each seed type remains largely unconstrained by observations \citep{miller15, trump15}.

Most quasar surveys, which observe $\gtrsim 1,000$ deg$^2$ down to $\sim20$ mag, are sensitive 
to only the most luminous quasar populations ($\sim10^{47}$ erg s$^{-1}$ in bolometric luminosity; $L_{\rm bol}$). 
These ultra-rare systems, which formed in biased regions of the early universe, place limited constraints on the BH seed population as they would have undergone 
sustained episodes of exponential growth, even for the most massive predicted seeds, thereby erasing 
the imprint of the initial seed mass distribution \citep[e.g.,][]{Tanaka_Haiman_2009,Volonteri_2010}.  
A complementary approach is to search for lower-luminosity quasars hosting SMBHs with masses closer 
to the predicted seed mass range at the earliest epochs possible 
\citep{somerville08,Valiante_2016,Ricarte_Natarajan_2018,Yung21,Li_2022}.  
Several deep optical surveys have attempted to do this by reaching a dex fainter in luminosity 
\citep[e.g.,][]{willott07, willott10a, matsuoka16, matsuoka22, kim18, kim20, fujimoto22}; however, these samples 
are still far more luminous than what is observed in the nearby universe 
($L_{\rm bol}\sim10^{43 - 44}$ erg s$^{-1}$; e.g., \citealt{gh07}, \citealt{liu18, liu19}), 
biasing our understanding of early SMBHs toward the most massive and active populations (however, see also \citealt{Mezcua18}).

Additional constraints on the seed mass distribution can be obtained by comparing the masses of 
high-redshift SMBHs to that of their host galaxies. 
In the local universe, well established scaling relationships exist between the mass of SMBHs and 
the bulge properties of their hosts (e.g., \citealp{Magorrian98, Gebhardt00, Ferrarese00, McConnell13, Sun15}).
However, offsets from this relationship at higher redshift can help constrain models of early BH growth and their co-evolution with galaxies \citep{Hirschmann10, Habouzit22, Hu_2022}. Observational studies have produced mixed results in this regard, with several reporting that SMBHs become increasingly overmassive relative to their hosts with increasing redshift (e.g., \citealp{Trakhtenbrot_Netzer10, Bennert11, Park15, Shimasaku19, Ding_2020, Neeleman21}), while other studies report no evolution in the local scaling relationship (e.g., \citealp{Willott17, Izumi19, Suh20}).
Pushing such studies to lower SMBH and host masses at high redshifts is expected to provide additional insight into the earliest seeds.  Not only are lower-luminosity AGN more representative of the normal BH population \citep{Habouzit22}, lower mass hosts have a relatively quiet merger history and so represent a robust “fossil record” of the initial BH-seed mass distribution \citep{Volonteri08, Volonteri09}.

JWST is expected to be a game changer on both fronts, allowing for the detection of lower luminosity quasars and the light of their host galaxies out to the epoch of cosmic reionization. Since its launch, JWST has already revealed the host morphologies of X-ray and optically selected AGN out to 
$z\sim 4$ \citep{Kocevski_2022,Ding_2022a}, detected the host light of a quasar at $z\simeq 6$ for the first time \citep{Ding_2022b}, and identified a candidate faint quasar at $z\simeq7.7$ \citep{Furtak22}.  Recently, \citet[][hearafter O23]{onoue23} reported a candidate low-luminosity AGN at $z\sim 5$ by exploiting the first NIRCam images of the CEERS program. This AGN candidate, \target, has a compact morphology and shows a rest-frame UV-to-optical spectral energy distribution (SED) that can be well explained by an unobscured quasar with $L_{\rm bol} = 2.5 \pm 0.3 \times 10^{44}$ erg s$^{-1}$ and strong Balmer and \OIII\ emission lines.
In addition, \cite{Carnall2023} recently reported the detection of broad H$\alpha$ emission from a quiescent galaxy at $z=4.658$ using JWST, from which they measure the central SMBH mass of $M_{\rm BH}=10^{8.7\pm0.1} M_\odot$.

Here we report on the detection of broad H$\alpha$ emission from two $z>5$ galaxies, including \target, using NIRSpec data obtained as part of the second epoch of CEERS observations.  The first source, CEERS 1670 at $z=5.242$, was identified as a result of targeted follow-up of CEERS-AGN-z5-1, while the second source, CEERS 3210 at $z=5.624$, was found serendipitously while inspecting the spectra of galaxies with photometric redshifts of $z>8$ in the literature.

We show that the SMBHs at the heart of these low-luminosity AGN have masses 1-2 dex lower than existing samples of luminous quasars with BH mass estimates at $z>5$.  We also examine the emission line ratios of both sources and place constraints on the relationship between SMBH and host mass in the lowest mass range yet probed in the early universe.  
Our analysis is presented as follows: in Section~\ref{sec:obs_data}, we describe the near-infrared imaging and spectroscopy used for this study, while in Section~\ref{sec:sample}, we discuss the properties of our sample.  In Section~\ref{sec:line_fit}, we outline our methodology for measuring the emission line properties of our sample. Section~\ref{sec:result} describes our results, and the implications of our findings are discussed in Section~\ref{sec:discussion}.  We use vacuum wavelengths for all emission-line features and, when necessary, the following cosmological parameters are used: 
$H_{0} = 70~{\rm km~s^{-1}~Mpc^{-1}}$, $\Omega_{\Lambda}=0.7$, and $\Omega_{\rm m}=0.3$.

\section{Observations \& Data Reduction} \label{sec:obs_data} 

The Cosmic Evolution Early Release Science Survey (CEERS) is an early release science program that covers 100 arcmin$^{2}$ of the Extended Groth Strip (EGS) with imaging and spectroscopy using coordinated, overlapping parallel observations by most of the JWST instrument suite (Finkelstein et al., in prep).  CEERS is based around a mosaic of 10 NIRCam pointings, with six NIRSpec and eight MIRI pointings observed in parallel.  Here we make use of NIRCam pointings 3 and 6 obtained on 21 June 2022, as well as NIRSpec pointing 4, obtained on 21 December 2022.  In each NIRCam pointing, data were obtained in the short-wavelength (SW) channel F115W, F150W, and F200W filters, and long-wavelength (LW) channel F277W, F356W, F410M, and F444W filters.  The total exposure time for pixels observed in all three dithers was typically 2835 s per filter.

The NIRSpec observations were taken with the G140M/F100LP, G235M/F170LP and G395M/F290LP $R\simeq 1000$ grating/filter pairs as well as with the $R\simeq 30-300$ prism, providing a complete coverage of the $1-5~\mu$m range with both configurations. The observation adopted a three-nod pattern, each of the nods consisting of a single integration of 14 groups (1036 s). The coadded spectra have a total exposure time of 3107 s in each spectral configuration. Targets for the microshutter array (MSA) configuration included sources selected using the NIRCam imaging in the field from CEERS epoch one (June 2022), especially prioritizing targets with photometric redshifts of $z>6$.  Each target was observed using a ``slitlet” aperture of three microshutters, and the design also included empty shutters for background subtraction.  The shutter configuration for observations taken with the medium resolution gratings and the prism are identical.

We performed an initial reduction of the NIRCam images in all four pointings, using version 1.5.3 of the JWST Calibration Pipeline\footnote{\url{http://jwst-pipeline.readthedocs.io/en/latest/}} with some custom modifications.  We used the current (15 July 2022) set of NIRCam reference files\footnote{\url{http://jwst-crds.stsci.edu}, \url{http://jwst\_nircam\_0214.imap}}, though  we note that the majority were created pre-flight, including the flats and
photometric calibration references. We describe our reduction steps in greater detail in \citet{finkelstein22} and \citet{bagley22}.  Coadding the reduced observations into a single mosaic was performed using the drizzle algorithm with an inverse variance map weighting \citep{fruchter02,Casertano00} via the Resample step in the pipeline. The output mosaics have pixel scales of 0\farcs03/pixel.

\begin{table}
\renewcommand\thetable{1} 
\caption{AGN Sample}\vspace{0mm}
\begin{center}
\vspace{-0.2in}
\begin{tabular}{ccccc}
\hline
\hline
Source Name & R.A. & Dec. & $z$ & $m_{\rm~F356W}$  \\
 & (deg) & (deg) & & (mag) \\
 \hline
CEERS~1670  &  214.823453 & 52.830281 & 5.242  & $25.8\pm0.01$ \\
CEERS~3210   &  214.809142 & 52.868484 & 5.624  & $26.9\pm0.04$ \\  
\hline 
\end{tabular}
\label{tab_sample}
\end{center}
 \tablecomments{CEERS~1670 is the same source as \target\ in \citetalias{onoue23}.}
\end{table}    


\begin{figure*}[t]
\centering
\epsscale{1.15}
\plotone{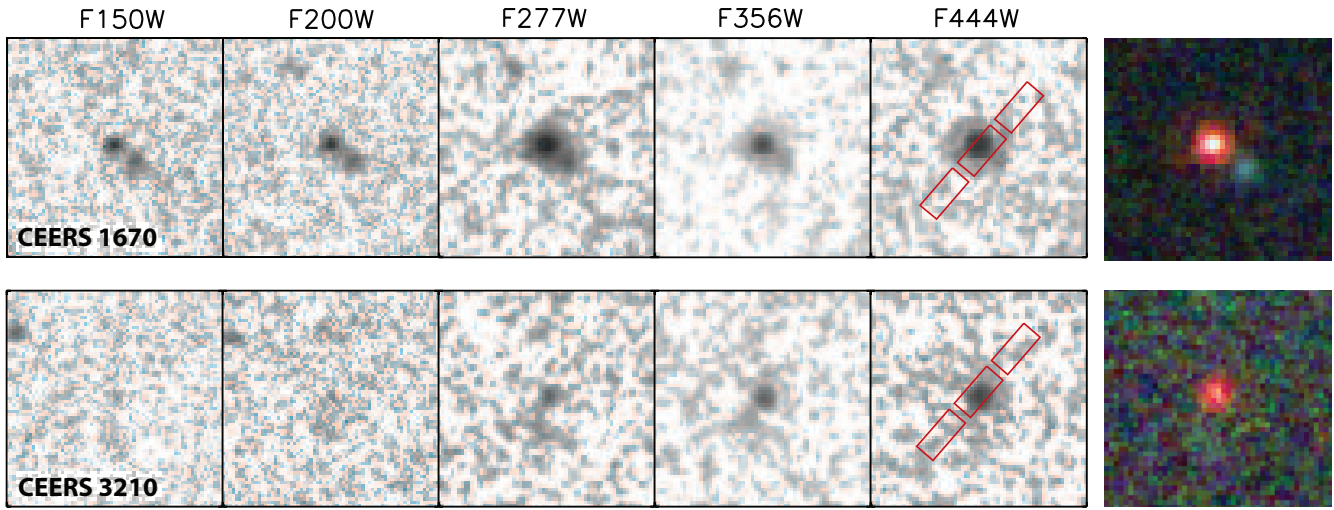}
\caption{JWST NIRCam images of our broad-line AGN sample at $z>5$ taken in the short-wavelength (F150W and F200W) and long-wavelength (F277W, F356W, and F444W) filters. The RGB images are composed of images in the F150W, F277W, and F444W filters. All images are $2^{\prime\prime}\times~2^{\prime\prime}$ in size.  The alignment of the NIRSpec microshutter aperture relative to each source is shown in red overtop the F444W image. \label{fig:thumbnails}}
\end{figure*}

Photometry was computed on PSF-matched images using SExtractor \citep{bertin96} v2.25.0 in two-image mode, with an inverse-variance weighted combination of the PSF-matched F277W and F356W images as the detection image.  Photometry was measured in all seven of the NIRCam bands observed by CEERS, as well as the F606W, F814W, F105W, F125W, F140W, and F160W HST bands using data obtained by the CANDELS and 3D-HST surveys \citep{grogin11, koekemoer11,brammer12,Momcheva16}. 

The CEERS NIRSpec observations (Arrabal Haro et al., in prep.) were reduced using version~1.8.5 of the JWST Science Calibration Pipeline with the Calibration Reference Data System (CRDS) mapping 1027, starting from the Level~0 uncalibrated data products (``\_uncal.fits'' files) available on MAST. Custom parameters were used for the \texttt{jump} step at the detector-level calibration for a better treatment of the ``snowballs''\footnote{\url{https://jwst-docs.stsci.edu/data-artifacts-and-features/snowballs-and-shower-artifacts}} produced by high-energy cosmic ray events, and a nodded background subtraction was adopted.

The reduced two-dimensional (2D) spectra (``s2d'') have a rectified trace with a flat slope. The current version (1.8.5) of the pipeline does not correctly identify source locations in the 2D spectra for one-dimensional (1D) spectra extraction. For the sources presented in this work, the 1D spectra were extracted using custom boxcar apertures centered on the visually identified continuum trace. Any remaining artifacts in the extracted spectra were masked after a detailed visual analysis.  The flux uncertainties of the reduced 1D spectra appear to be underestimated by a factor of $\sim$2, as estimated from the normalized median absolute deviation (NMAD) of the flux in line-free regions, and so we rescale the flux uncertainty of each spectrum by a factor equal to the ratio of the line-free NMAD to the median pipeline uncertainty.

The current (version~1.8.5) of the NIRSpec MSA data reduction uses a flux calibration that relies on pre-flight knowledge of the instrument, which is known to differ from the post-launch performance (see Figure~20 of \citealt{rigby22}). The pipeline applies a correction for ``slit losses’’ outside the MSA aperture using a {\it pathloss} reference file based on a pre-launch model for point sources that has not yet been fully verified on orbit.  This correction may be inaccurate for extended sources or non-default spectral extraction apertures, and indeed by comparing spectroscopic fluxes to NIRCam photometry we find some evidence that further corrections are required (see, e.g., section 3).  While this may impact our interpretation of individual line fluxes or luminosities,
the \textit{relative} spectrophotometry of the reduced spectra is measured to be reliable, with line ratios of doublets (\OIII~$\lambda\lambda$4960, 5008; \citealt{storey00}) and Balmer lines \citep{osterbrock89} that match physical expectations (see additional discussion in \citealt{trump23}, Arrabal Haro et al., in prep.).

\begin{figure*}[t]
\centering
\epsscale{1.1}
\plotone{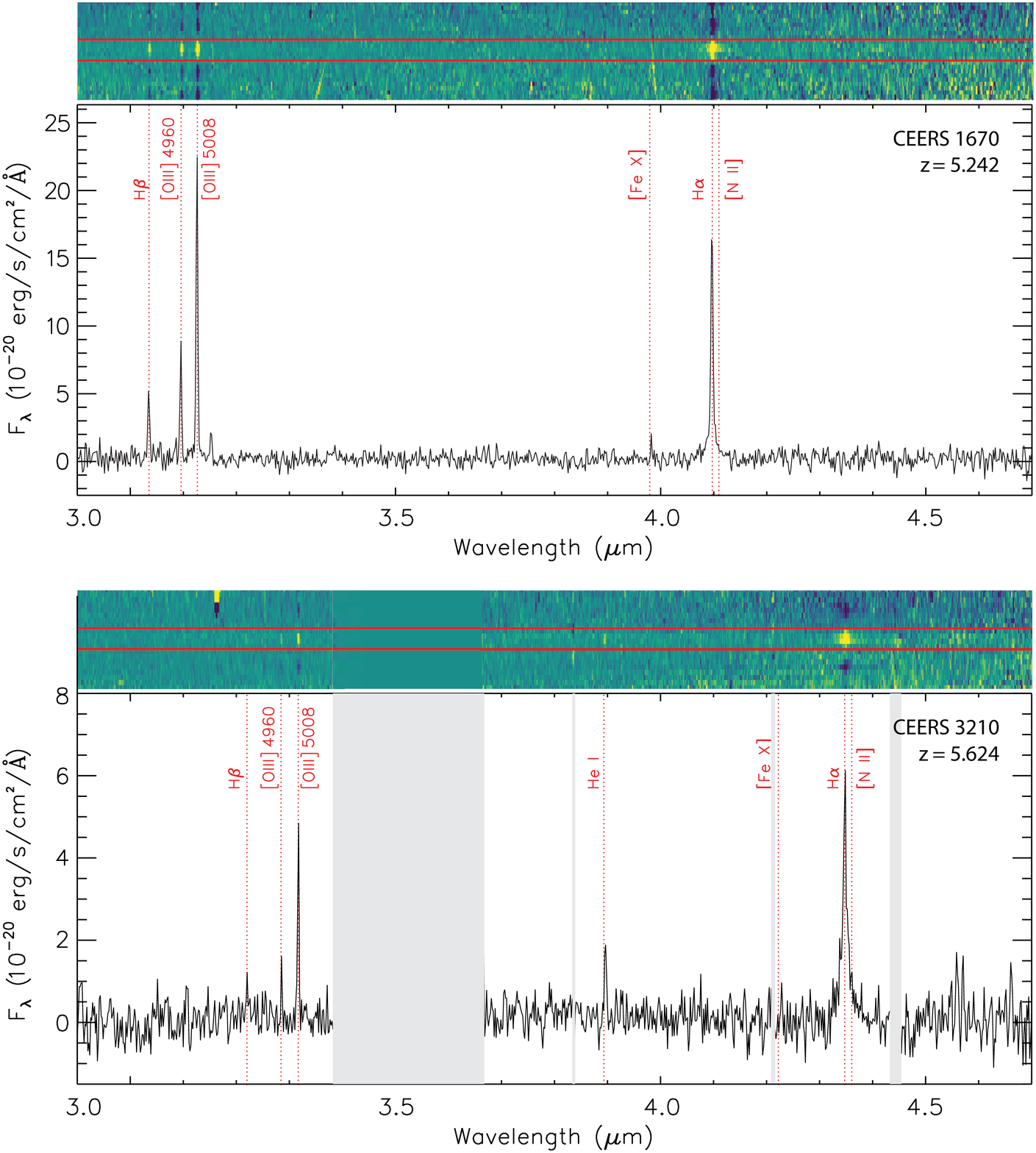}
\caption{NIRSpec spectra of sources CEERS~1670 and CEERS~3210 taken in the G395M grating with $R\sim1000$.  The 2D spectra are shown above with extraction windows highlighted in red.  Grey regions in both the 1D and 2D spectra indicate regions masked due to artifacts identified via visual inspection.  The location of several prominent emission lines are noted.
\label{fig:spectra}}
\end{figure*}

\begin{figure*}
\centering
\includegraphics[width=\linewidth]{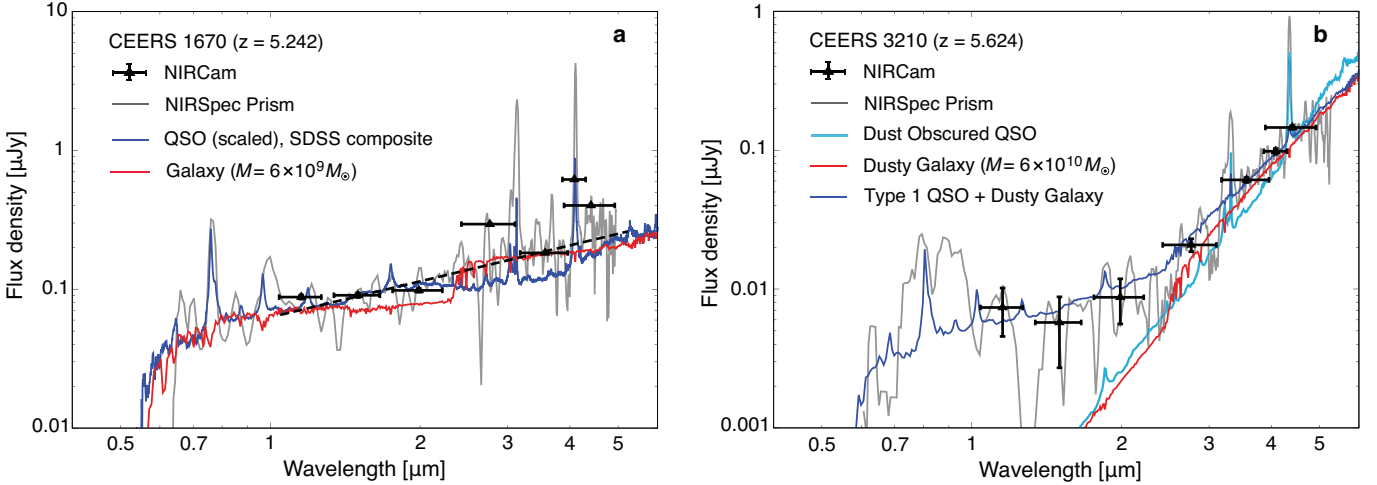}
\caption{
The SEDs of the two low-luminosity AGN (CEERS~1670 and CEERS~3210) obtained with the JSWT NIRSpec and NIRCam.
Left panel (a): the continuum spectral shape is explained by the composite quasar spectrum of \citetalias{VB_2001} 
scaled to match the photometry of CEERS~1670 (blue), and is fitted well with a single power law with an index of $\alpha_\lambda=-1.14$ (dashed).
The galaxy SED model with $M_\star \simeq 6.0\times 10^9~\msun$ is overlaid (red),
where the stellar continuum in the F356W filter becomes comparable to the observed F356W flux density.  This gives a robust upper bound of the underlying stellar population. Right panel (b): the source has a blue continuum spectrum with a UV slope of $\alpha_\lambda < -3.0$ at $\lambda_{\rm obs}\simeq 1-2~\mu$m and a very steep continuum spectrum ($\alpha_\lambda \simeq 2.0$).
The redder part can be explained either by a heavily obscured quasar (cyan) or a dusty starburst galaxy (red).
As a possible explanation of the blue excess in the spectrum, the unobscured broad-line AGN contribution is added to the dusty starburst galaxy (blue).
In the dusty galaxy model, the stellar mass is set to $M_\star \simeq 6\times 10^{10}~\msun$
(see the text in Section~\ref{sec:SED3210}).
}\hspace{5mm}
\label{fig:SED}
\end{figure*}

\begin{figure*}[t]
\centering
\epsscale{1.15}
\plotone{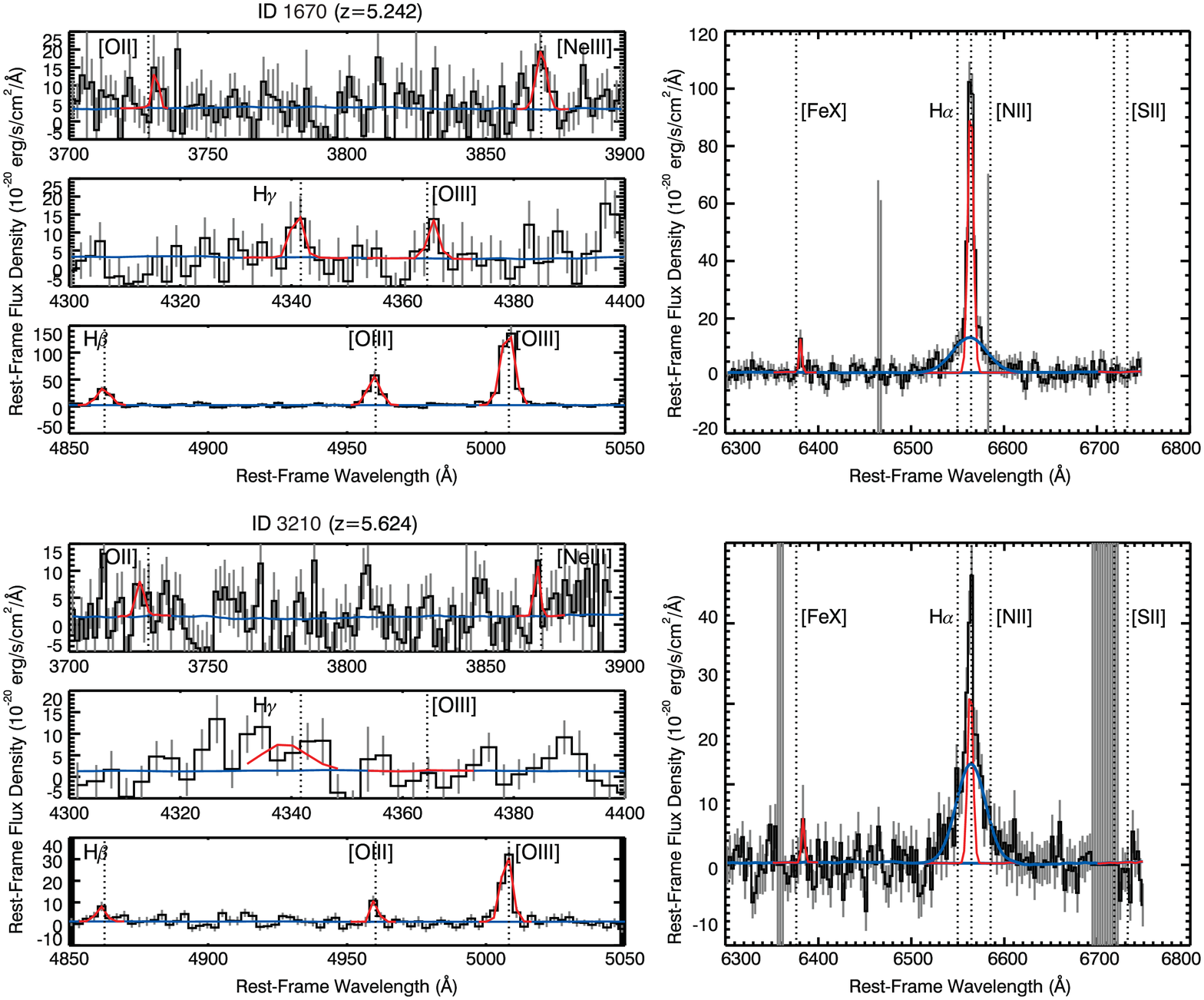}
\caption{The rest-frame spectra (black histograms) and associated uncertainty (gray error bars) of both sources in regions with emission-line features. Red lines show the best-fit Gaussians for narrow emission lines and the blue line shows the best-fit broad component for \Ha, which have a FWHM of $2060\pm286$ and $1802\pm204$ km s$^{-1}$ for CEERS~1670 and CEERS~3210, respectively.
\label{fig:line_fits}}
\end{figure*}

\section{Sample Description}\label{sec:sample}

During the initial inspection of our reduced NIRSpec data, we identified two sources with broad H$\alpha$ emission.  Information on these sources, referred to as CEERS~1670 and CEERS~3210, is listed in Table \ref{tab_sample}.  CEERS~1670 was observed as a result of targeted follow-up of the AGN candidate \target ~identified by \citetalias{onoue23}. CEERS~3210 was selected for observation as it was previously identified as a candidate massive galaxy at $z=8.13$ by \citet{labbe22} and a potential strong-line emitter at $z=5.72$ by \citet{Perez-Gonzalez22}. NIRCam images of both sources are shown in Figure~\ref{fig:thumbnails}, while their 1D and 2D spectra from the G395M grating are shown in Figure~\ref{fig:spectra}.  Our derived redshifts, based on the \OIII~$\lambda\lambda$4960, 5008 narrow lines, for CEERS~1670 and CEERS~3210 are $z=5.242$ and $z=5.624$, respectively.

Neither source is directly detected in the deep (800 ksec) Chandra X-ray observations of the CEERS field from the AEGIS-XD survey \citep{nandra15}.  However, the shape of their SEDs, coupled with the existence of broad-line emission in their spectra, suggest both sources host low-luminosity AGN.

In Figure~\ref{fig:SED}, we show the NIRCam photometry and NIRSpec prism spectrum of both CEERS~1670 and CEERS~3210.  In the case of CEERS~1670, we find the prism spectrum must be scaled by a factor of $2\times$ to match the NIRCam broad-band photometry.  This may be due to potential slit losses, as CEERS~1670 sits near the edge of its microshutter slit, the outline of which can be seen in Figure \ref{fig:thumbnails}.  We find no such correction is needed for the CEERS~3210 prism spectrum.

As discussed by \citetalias{onoue23}, the broad-band photometry of CEERS~1670 is well reproduced by a continuum model with a single power-law function, with the exception of filters that are affected by strong line emission, namely F277W, F410M, and F444M.  
A single power-law fit to the other four filters yields the best-fit power-law slope $\alpha_\lambda = -1.14\pm0.03$ ($\equiv {\rm d~ln}~F_{\lambda}/{\rm d~ln}~\lambda$),  which is consistent with a typical value for unobscured quasars (e.g., \citealp{fan01, VB_2001}).  
This power-law model yields the absolute magnitude at rest-frame 1450 {\rm \AA} of $M_{1450}=-19.44 \pm 0.05$ mag.
Likewise, the monochromatic luminosity at rest-frame 3000 {\rm \AA} and 5100 {\rm \AA} is $L_{3000} = (4.83 \pm 0.09) \times 10^{43}$ erg s$^{-1}$ and $L_{5100} = (4.48 \pm 0.08) \times 10^{43}$ erg s$^{-1}$, respectively.
We find that a low-redshift composite spectrum of quasars (the blue model in Figure \ref{fig:SED}a) from \citet[hereafter VB01]{VB_2001} scaled to match the photometry can explain the observed spectral shape of CEERS~1670 well.

The SED of CEERS~3210 shows more complexity.  The source has a blue continuum spectrum with a UV slope of $\alpha_\lambda =-3.0 \pm 0.3$ at $\lambda_{\rm obs}\simeq 1-2~\mu$m and a very steep continuum spectrum ($\alpha_\lambda = 1.8 \pm 0.2$) with strong Balmer and {\OIII} emission lines at longer wavelengths.
This steep spectral slope, coupled with the broad \Ha~emission we detect, suggests that this source is a heavily obscured, broad-line AGN \citep[e.g.,][]{Gregg_2002}.
In Figure~\ref{fig:SED}b, we overlay the composite SED of low-redshift broad-line AGN (\citetalias{VB_2001}) reddened assuming a color excess of $E(B-V)=0.9$ and the extinction law discussed in \citet{calzetti00}. Note that this model shown with the cyan curve is essentially the same as the QSO2 SED template provided in \citet{Polletta_2006}.  This model traces the observed prism continuum at $\lambda_{\rm obs}\gtrsim 3~\mu$m well; however, the obscured broad-line AGN model does not explain the blue side of the observed spectrum, requiring additional components at these shorter wavelengths.  We discuss more complex SED models, including fits using hybrid galaxy plus AGN models, in Section~\ref{sec:SED3210}.

\section{Line Fitting Analysis}\label{sec:line_fit}
 
The NIRSpec spectra of CEERS~1670 and CEERS~3210 include several prominent emission lines. The G395M/F290LP spectrum of both sources includes strong H$\alpha$, H$\beta$, and \OIII~$\lambda\lambda$4960, 5008 emission, and CEERS~3210 also features a He I~$\lambda5877.25$ line. Both sources exhibit a weak line near the expected wavelength of the \FeX~$\lambda$6376 coronal emission line.  The G235M/F170LP spectrum of both sources includes the \NeIII~$\lambda3870.86$ line, while CEERS~3210 also exhibits the \Hg~$\lambda4341.69$ and auroral \OIII~$\lambda4364.44$ lines.

We measure line fluxes and uncertainties with a Levenberg-Marquardt least-squares method implemented by the \texttt{mpfit} IDL code\footnote{https://pages.physics.wisc.edu/$\sim$craigm/idl/fitting.html}. We fit isolated lines with single Gaussians and simultaneously fit multiple Gaussians for features in the Balmer line regions. The results of our line fits are shown in Figure \ref{fig:line_fits}.

To account for potential broad components, we fit the \Ha\ line with two Gaussians: one narrow with width $\sigma<350$~km~s$^{-1}$ and one broad with width $\sigma>350$~km~s$^{-1}$. We also attempted to include additional Gaussian components for the \NII~$\lambda\lambda$6550, 6585 doublet, constraining the line widths and relative line centers to narrow \Ha, but found that the \NII\ lines are not significantly ($>$3$\sigma$) detected and their inclusion does not improve the $\chi^2_0$ of the fit. We report the $1\sigma$ upper limit for \NII~$\lambda$6585 but do not include it in the fits for broad and narrow \Ha.

\begin{figure*}[t]
\centering
\epsscale{1.15}
\plotone{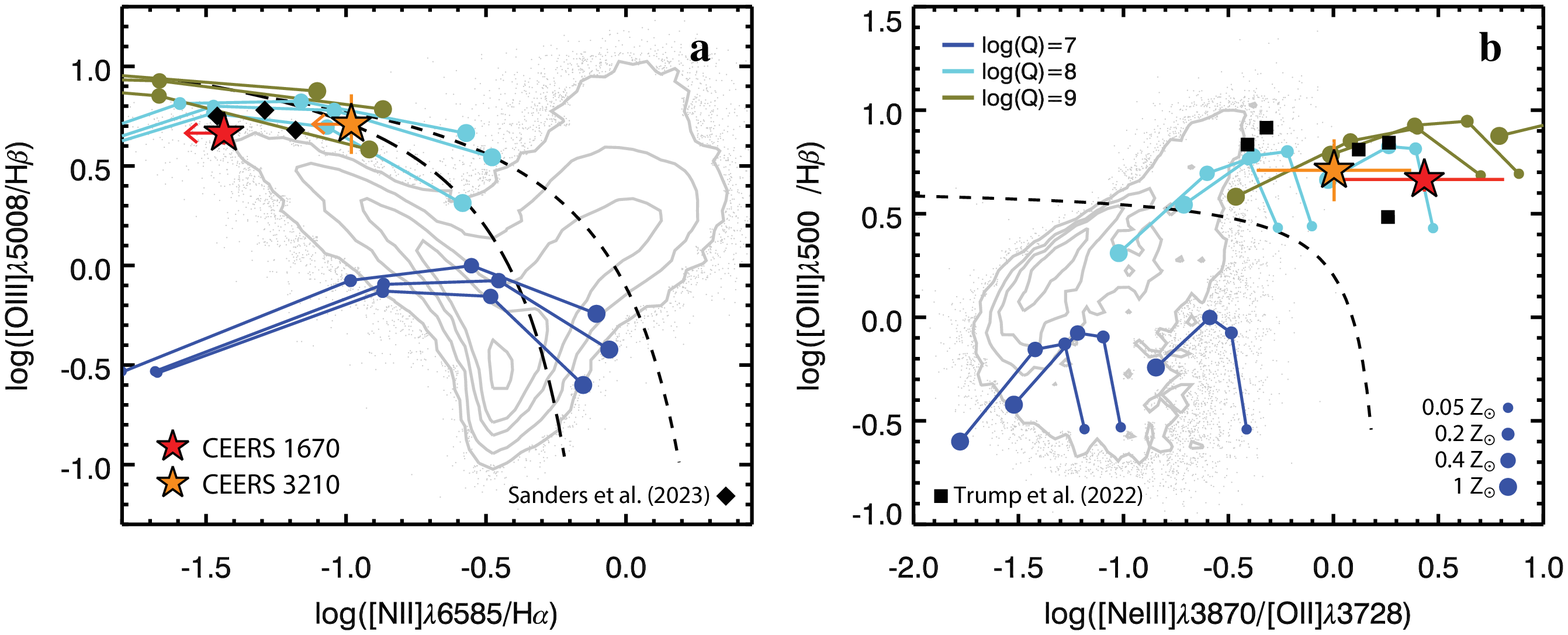}
\caption{Left Panel (a): The BPT emission-line diagnostic diagram.  The gray contours denote the distribution of local star-forming galaxies and AGN as measured by the SDSS survey \citep{sdss}. Black diamonds denote stacked line ratios of CEERS galaxies at $z \sim 5.6$, $z \sim 4.5$, and $z \sim 3.3$ \citep{Sanders_2023}. The black long and short-dashed lines denote the $z=0$ and $z=2.3$ boundary between the star-forming and AGN regions of the diagram defined by \citet{Kauffmann_2003} and \citet{Kewley_2013a}, respectively. Right Panel (b): the OHNO diagnostic diagram. Black squares denote line ratios of SMACS ERO galaxies at $5.3<z<8.5$ \citep{trump23} and gray contours denote the distribution of $z \sim 0$ SDSS galaxies. The dashed line denotes the boundary between star-forming and AGN regions as defined in \citet{Backhaus22}. Colored curves in both panels show MAPPINGS~V photoionization models \citep{Kewley_2019}. The three color-coded sets of curves and points along those curves correspond to different ionization parameters and metallicities, as indicated by the legends, with three curves for each color corresponding to different gas pressures as described in the text.  Both of our $z \sim 5$ AGN have narrow-line ratios that are consistent with low metallicity and high ionization, with little difference from the emission-line ratios observed for other populations of high-redshift galaxies.
\label{fig:BPT}}
\end{figure*}

We also performed a simultaneous fit of the \Hb\ emission-line region with components for narrow \Hb\ and the \OIII~$\lambda\lambda$4960,5008 doublet. In both systems we tested a fit that included an additional broad ($\sigma>350$~km~s$^{-1}$) \Hb\ component but found that this component is only marginally ($<$1$\sigma$) detected and including it increases the $\chi^2_0$ of the fit. We report 1$\sigma$ upper limits for putative broad \Hb\ emission that assume the same width as the broad \Ha\ component applied to the local noise of the \Hb\ region.

Finally, we fit single narrow Gaussians for the \OII~$\lambda3728.48$ (the 3727+3729 doublet is blended in the $R \simeq 1000$ medium-resolution NIRSpec grating), \NeIII~$\lambda3870.86$, \Hg~$\lambda4341.69$, and \OIII~$\lambda4364.44$. The \NeIII\ line is significantly ($>$3$\sigma$) detected in CEERS~1670 and all the other lines are only marginally ($<$3$\sigma$) detected.

\section{Results}\label{sec:result}

\subsection{Emission Line Properties} \label{sec: line_properties}  

Our two AGN are identified from their broad \Ha\ emission. As described above, we use a two-component fit with both narrow and broad Gaussian components in which the line centers, widths, and fluxes are free parameters. These broad+narrow fits have significantly lower $\chi_0^2$ than single-Gaussian fits for the \Ha\ lines. Both objects have best-fit narrow \Ha\ components that are unresolved in the $R \sim 1000$ NIRSpec spectra, with narrow-\Ha\ widths of $\sigma = 135 \pm 9$~km~s$^{-1}$ and $\sigma = 131 \pm 24$~km~s$^{-1}$ for CEERS~1670 and CEERS~3210, respectively. The best-fit broad \Ha\ components have $\sigma = 840\pm120$~km~s$^{-1}$ and ${\rm FWHM} = 2060\pm290$~km~s$^{-1}$ for CEERS~1670 and $\sigma = 720\pm87$~km~s$^{-1}$ and ${\rm FWHM} = 1800\pm200$~km~s$^{-1}$ for CEERS~3210 (fitting $\sigma$ and FWHM independently).

In contrast, the \Hb\ emission lines of both objects are best-fit by single narrow Gaussians, with no statistical improvement from including a broad component. Both \Hb\ lines appear to be unresolved, with best-fit single-Gaussian widths of $\sigma = 145\pm17$~km~s$^{-1}$ for CEERS~1670 and $\sigma = 108\pm33$~km~s$^{-1}$ for CEERS~3210. We compute upper limits for a potential (undetected) broad \Hb\ component by assuming a Gaussian of the same width as the measured \Ha\ broad lines with the noise properties of the \Hb\ region in the spectrum. In both cases the upper limit for potential \Hb\ broad emission is statistically consistent with a broad $\Ha/\Hb = 3.1$ \citep{osterbrock89}: CEERS~1670 has a lower limit of broad $\Ha/\Hb > 2.4$ (3$\sigma$) and CEERS~3210 has a lower limit of $\Ha/\Hb > 3.0$ (3$\sigma$). In other words, both CEERS~1670 and CEERS~3210 are consistent with (undetected) broad \Hb\ emission that matches typical Type 1 AGN \Ha/\Hb\ ratios, and the lack of observed broad \Hb\ in CEERS~1670 and CEERS~3210 cannot be used to classify them as intrinsic Type 1.5 AGN.

The narrow Balmer emission lines imply modest dust attenuation in both objects. CEERS~1670 has a measured narrow-line Balmer decrement of $\Ha/\Hb = 3.9\pm0.5$ and CEERS~1670 has a narrow-line $\Ha/\Hb = 5.3\pm2.1$. We use these Balmer decrements as priors to inform the SED fitting in Section~\ref{sec:SED1670} and \ref{sec:SED3210}.

Intriguingly, both AGN have weak emission-line features that are consistent with marginally-detected \FeX~$\lambda$6376, as seen in Figures \ref{fig:spectra} and \ref{fig:line_fits}. \FeX\ is a coronal emission line with an ionization potential of 262~eV that is observed in low-mass AGN in the local universe \citep[e.g.,][]{molina21}. The putative \FeX~emission lines are marginally detected with SNR=2.4 for CEERS~1670 and only SNR=1.5 for CEERS~3210. Both lines are best-fit to be slightly redder than the other narrow-line features: if the marginal detections represent genuine emission lines then they may indicate a kinematic offset between the extreme-ionization coronal gas and the narrow-line region.

Finally, in Figure~\ref{fig:BPT} we plot the narrow emission-line ratios of both sources on the BPT (\OIII/\Hb\ versus \NII/\Ha; \citealt{Baldwin81}) and OHNO (\NeIII/\OII\ versus \OIII/\Hb; \citealt{Backhaus22}) line-ratio diagnostics that are commonly used to classify galaxies as dominated by emission from AGN or star formation. The colored curves in Figure~\ref{fig:BPT} indicate MAPPINGS~V photoionization models from \citet{Kewley_2019}, with different colored curves for different ionization ($\log(Q/[{\rm cm~s}^{-1}])=[7,8,9]$ increasing left to right), metallicity along each curve ($Z/Z_\odot=[1,0.4,0.2,0.05]$ as indicated in the legend), and curves shown for each of three thermal pressures ($\log (P k_{\rm B}^{-1}/[{\rm K~cm}^{-3}])=[7,8,9]$). The MAPPINGS~V models use $\alpha$-enhanced abundances as described in \citet{nicholls17}, such that low metallicities include enhanced relative abundances of O and Ne (and a lower relative abundance of N). Figure~\ref{fig:BPT} also includes comparison samples of high-redshift galaxy line ratios from early JWST spectroscopy: stacked CEERS measurements from \citet{Sanders_2023} in the BPT and SMACS~ERO galaxies from \citet{trump23} in the OHNO diagram.

At low redshift ($z \lesssim 2$), AGN typically have higher \NII/\Ha, \OIII/\Hb, and \NeIII/\OII\ ratios due to harder ionizing radiation from the AGN accretion disk, and line-ratio diagnostics shown in Figure \ref{fig:BPT} can be used to separate AGN from star-forming galaxies. However, high-redshift galaxies show systematic offsets relative to galaxies and AGN at $z=0$, with higher ionization and lower metallicity in both AGN and from star-forming \HII\ regions \citep{Shapley_2005,Erb_2006,Liu_2008,Kewley_2013a,Kewley_2013b,Sanders_2023}. Both CEERS~1670 and CEERS~3210 have high \OIII/\Hb, low \NII/\Ha, and high \NeIII/\OII\ line ratios that are consistent with MAPPINGS~V photoionization models for high ionization ($\log (Q/[{\rm cm~s}^{-1}]) \simeq 8$) and moderately low metallicity ($Z/Z_\odot \simeq 0.2-0.4$). 

The AGN line ratios and interstellar medium conditions implied in Figure~\ref{fig:BPT} are virtually indistinguishable from star-forming galaxies observed at similar redshifts, since high-redshift \HII\ regions have similarly high ionization and low metallicity to these $z \sim 5$ AGN narrow-line regions.  Photoionization models show that low-metallicity AGN can have similar \OIII/\Hb~and \NII/\Ha~ratios and lie within or even below the star-forming branch \citep{Groves_2006, Feltre16}. Although low-metallicity AGN are rare in the local universe \citep[e.g.,][]{Storchi-Bergmann98, Groves_2006}, recent simulations that make use of the AGN photoionization models presented in \citet{Feltre16} predict that high-redshift, low-metallicity AGN should primarily occupy the top portion of the local star-forming branch \citep{Hirschmann19, Hirschmann22}, in agreement with our findings.  
The fact that neither source is X-ray detected and that their BPT line ratios are similar to that of star-forming galaxies observed at the same redshift means that their broad-line emission may be one of the few ways to detect these high-redshift low-luminosity AGN.
Other possible approaches include diagnostics with high-ionziation and extreme-ionization lines (e.g., \HeII\ and \NeV; \citealt{Feltre16, Nakajima_2022, Cleri23}).
Preselection with photometric colors may also be useful to select fast-growing BHs with $M_{\rm BH} \sim 10^{6-7}~\msun$ 
in metal-poor environments \citep{IO_2022}.

\begin{table*}
\renewcommand\thetable{2} 
\caption{Derived AGN Properties}\vspace{0mm}
\begin{center}
\begin{tabular}{lcccccccc}
\hline
\hline
ID  & $M_{1450}$ & $L_{5100}$ & $L_{\rm H \alpha}$(broad)  & FWHM$_{\rm H\alpha, broad}$ & $M_{\rm BH}$ &
$\lambda_{\rm Edd}$ & 
$M_\star$ & $({\rm H \alpha/H\beta})_{\rm obs}$ \\
 & (mag) & ($10^{43} {\rm erg~s^{-1}}$) & ($10^{42} {\rm erg~s^{-1}}$) & (${\rm km~s^{-1}}$) & ($10^{7}~\msun$)& & ($10^{9}~\msun$)\\
\hline 
1670   & $-19.4\pm0.05$ & $4.48\pm0.08 $ & $1.64\pm0.21$ & $2060\pm290$ & $ 1.3\pm0.4 $ & $0.15\pm0.04$ & $<6.0 $ & $3.9\pm 0.5$ \\\hline
3210  & \multicolumn{2}{c}{See text} &  $1.67\pm0.16 $ & $1800\pm200$ & $0.90\pm0.22$ & $0.29\pm0.08$ & $<60.0$ & $5.3\pm 2.1$ \\
3210$_{\hspace{0.01in}A_{v}=\hspace{0.01in}4}$   &  \multicolumn{2}{c}{See text} &  $34.4\pm3.4  $ & $1800\pm200$ & $4.7\pm1.2$ & $3.5\pm0.9$ & $<60.0$ & $5.3\pm 2.1$ \\
\hline 
\end{tabular}
\label{tab:reesult}
\end{center}
\tablecomments{The BH mass for CEERS 1670 uses  $L_{5100}$ estimated from the photometric SED and the line width of broad H$\alpha$ (FWHM$_{\rm H\alpha, broad}$) (Equation~\ref{eq: Kaspi00}), while for CEERS 3210 we use FWHM$_{\rm H\alpha, broad}$ and line luminosity of broad H$\alpha$  (Equation~\ref{eq: GH05}). The bolometric luminosity is also converted from $L_{\rm H\alpha}$ for CEERS 3210.
In the third row, we show the case when CEERS 3210 is heavily dust-reddened with $A_V=4$.  The H$\alpha$ luminosities are reported as observed, with no correction for potential slit losses.}
\end{table*}

\subsection{Virial BH Mass Estimates} \label{sec: MBH} 

In this section, we estimate the virial BH masses of the two broad-line AGN assuming that their broad H$\alpha$ emission traces the kinematics of gas in the broad-line-region.
The single-epoch BH mass estimation method is best calibrated against the width of the broad H$\beta$ emission line and the rest-frame 5100 \AA\ continuum luminosity ($L_{5100}$) using the reverberation mapping technique \citep[e.g.,][]{Kaspi00}.
However, since we do not detect a broad H$\beta$ component in our spectra, we instead employ the BH mass relationship proposed by \citet[hereafter GH05]{greene_ho05}, which relies entirely on H$\alpha$ emission.  This method has been widely used in, for example, BH mass estimates for AGN in dwarf galaxies \citep[e.g.,][]{reines13, baldassare15}.  This recipe is based on empirical correlations between Balmer emission-line luminosities and $L_{5100}$ and between the line widths of H$\beta$ and H$\alpha$.

In terms of the broad H$\alpha$ line width and $L_{5100}$, the BH mass formula is expressed as:
\begin{equation} \label{eq: Kaspi00}
M_{\rm BH} = 5.04 \times 10^6~\msun \left( \frac{L_{5100}}{10^{44}\ {\rm erg\ s^{-1}}}\right)^{0.64}  \left(\frac{{\rm FWHM_{\rm H\alpha}}}{10^3\ {\rm km\ s^{-1}}} \right)^{2.06}.
\end{equation} 
This equation is based on the formula of \citet{Kaspi00} for H$\beta$ with the H$\beta$ line width substituted with that of H$\alpha$ \citepalias[Equation~3 of ][]{greene_ho05}.  It is important to note that this equation assumes that the 5100 \AA\ continuum luminosity is dominated by light from the AGN.  Alternatively, we can directly apply the virial BH mass recipe of \citetalias{greene_ho05}, which is based on the broad H$\alpha$ line width and luminosity:
 \begin{equation} \label{eq: GH05}
 M_{\rm BH} = 2.0 \times 10^6 \left( \frac{L_{\rm H\alpha}}{10^{42}\ {\rm erg\ s^{-1}}}\right)^{0.55}  \left(\frac{{\rm FWHM_{\rm H\alpha}}}{10^3\ {\rm km\ s^{-1}}} \right)^{2.06} M_\odot.
 \end{equation} 

First, we use the line width of the broad H$\alpha$ component detected in our NIRSpec spectroscopy,
corrected for the $R \sim 1000$ instrumental resolution,
and $L_{5100}$ derived from the photometric SEDs to estimate the virial BH masses of CEERS~1670.  Using Equation 1 results in a BH mass of $M_{\rm BH} = 1.3\pm 0.4 \times 10^7\ M_\odot$, with the Eddington ratio of $L_{\rm bol}/L_{\rm Edd} = 0.15\pm0.04$.
We use the bolometric luminosity inferred from $L_{3000}$ to be consistent with other $z>5$ BH mass estimates in the literature. 
We apply a bolometric correction of $L_{\rm bol} = 5.15 L_{3000}$ \citep{Richards06} to derive $L_{\rm bol}=2.49 \pm 0.04 \times 10^{44}$ erg s$^{-1}$.  
Using instead the H$\alpha$ line width and luminosity, Equation 2 yields $M_{\rm BH} = 1.1\pm 0.3 \times 10^7\ M_\odot$. This value is more systematically uncertain than our first estimate, although consistent within the 1$\sigma$ error, owing to potential slit losses (see Section~\ref{sec:obs_data}).

The BH mass estimate for CEERS~3210 is complicated because of its potentially obscured nature. Taking the observed H$\alpha$ luminosity at face value and applying Equation~\ref{eq: GH05}, we derive a mass of $M_{\rm BH} = 9.0\pm 2.2 \times 10^6\ M_\odot$.
We caution that this value is likely a lower limit since the H$\alpha$ emission is likely affected by dust extinction.  
If we assume that a dust-reddened AGN continuum dominates the observed rest-optical spectrum with $A_V=4$ (see Section~\ref{sec:SED3210}), the inferred BH mass could be as high as $M_{\rm BH} = 4.7\pm 1.2 \times 10^7\ M_\odot$.
A careful decomposition of the AGN/host components, and, if the AGN is dust-reddened, measurements of AGN continuum luminosity at rest-frame infrared wavelengths \citep{greene14, kim15} are required to better estimate the intrinsic continuum luminosity and subsequently the virial mass for this AGN.

\section{Discussion}\label{sec:discussion}

\subsection{The $M_{\rm BH}-L_{\rm bol}$ Distribution}\label{sec:ML}
\begin{figure}
\centering
\includegraphics[width=\linewidth]{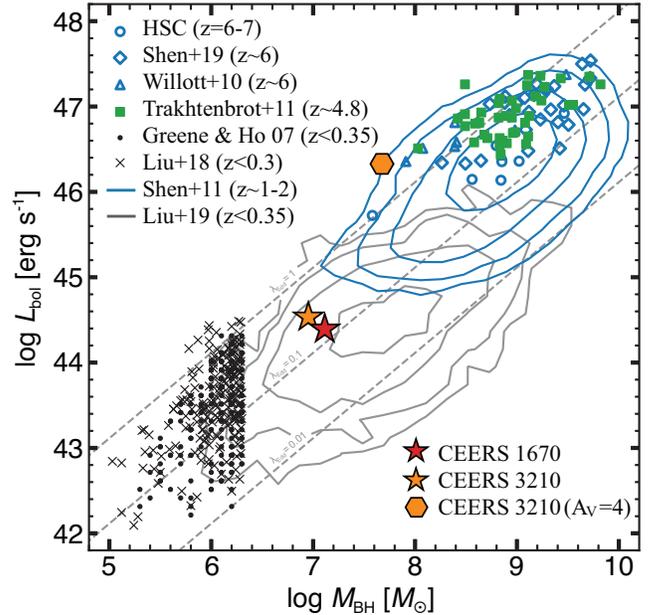}
\caption{
The BH mass - bolometric luminosity plane.  Quasar samples at $z\geq5$ are shown as blue and green symbols and contours, while low redshift AGN are shown in black.  CEERS~1670 and CEERS~3210 have BH masses 1-2 dex below that of known high redshift quasars and more comparable to those of typical nearby AGN. 
} \label{fig:MBH}
\end{figure}

The successful spectroscopic identification of two low-luminosity broad-line AGN at $z>5$ opens up a new parameter space for high-redshift AGN studies, thanks to the unprecedented infrared sensitivity of JWST and the multiwavelength photometric dataset available in the EGS field. Figure~\ref{fig:MBH} shows the distribution of $z\gtrsim5$ AGN in the BH mass - bolometric luminosity plane with the two new low-luminosity AGN shown in red and orange. 

As is discussed in \citetalias{onoue23}, CEERS~1670 is 2--3 dex fainter than known quasars at $z\gtrsim5$ \citep[e.g.,][]{willott10, trakhtenbrot11, shen19, onoue19, matsuoka19, kato20} and more comparable to those of typical nearby AGN \citep[e.g.,][]{liu19}.
The virial BH mass estimate we present above now shows that this low-luminosity AGN is by far the least-massive BH known in the universe at the end of cosmic reionization. 
The modest Eddington ratio of CEERS~1670 suggests that this AGN has been identified after its rapid accretion mode has ended, although it is possible the system will experience future bursts of heavy accretion \citep{Li_2022}.

For CEERS~3210, if we use the observed H$\alpha$ luminosity without an extinction correction, then the BH powering this AGN may be comparably low-mass as CEERS~1670.  
However, if we assume heavy dust attenuation ($A_V=4$), it becomes a BH accreting at a rate above the Eddington limit.  In Figure~\ref{fig:MBH}, we show our results assuming both no extinction for the H$\alpha$ luminosity and $A_{V} = 4$ with the bolometric luminosity converted from $L_{5100}$ estimated from the H$_{\alpha}$ luminosity.
Adopting a more moderate level of dust extinction inferred from the observed Balmer decrement in the NIRSpec spectrum (${\rm H}\alpha/{\rm H}\beta=5.3$; $A_V=1.9$), brings the bolometric luminosity of the source closer to the Eddington value.  
Thus, CEERS~3210 is likely in its most active mode of accretion and on the way to expelling the material that currently obscures it. 
\citet{fujimoto22} report a dust-reddened AGN at $z=7.19$, the BH mass of which is estimated to be $M_{\rm BH}\lesssim 10^8~\msun$ based on the upper limit of its X-ray luminosity. Although not confirmed, their AGN and CEERS~3210 may be drawn from the same population of high-redshift dust-reddened AGN.  We discuss this scenario in greater detail in Section~\ref{sec:SED3210} below.

\subsection{Constraints on the Host Galaxy Mass of CEERS 1670}\label{sec:SED1670}

Figure~\ref{fig:SED}a shows the prism spectrum and NIRCam photometric flux densities of CEERS~1670.
As discussed in Section~\ref{sec:sample}, the continuum spectral shape can be explained by the low-redshift composite 
quasar spectrum of \citetalias{VB_2001}.  
Since the observed spectrum is dominated by the central AGN contribution,
it is challenging to estimate the stellar mass of the host galaxy in a plausible way.
\citetalias{onoue23} conducted the SED fitting analysis for the photometric data
using templates of metal-poor galaxies \citep{Inoue_2011}.
The best-fit model with pure galaxy SEDs, where the quasar contribution is neglected, suggests 
a case with metallicity $Z= 0.2~\zsun$, stellar age $500$ Myr, star formation rate (SFR) $3.6~\msun~{\rm yr}^{-1}$,
whose stellar mass is $1.8\times 10^9~\msun$.
This value is considered to be an upper bound of the stellar mass among the SED templates \citetalias{onoue23} explored,
but the true upper bound depends sensitively on the properties of the assumed stellar population.
In the following, we give a robust  upper bound of the stellar mass built up in the host galaxy at $z\gtrsim 5$,
assuming the SED model parameters that yield a high mass for the given stellar luminosity.

One advantage of focusing on $z>5$ galaxies is that the stellar age is limited to the age of the Universe, e.g., $t\simeq 1$ Gyr at $z=5.7$. Although the star formation history (SFH) in the galaxy is unconstrained, the mass-to-light ratio ($M_\star/L_\star$) in the rest-frame optical and near-infrared band tends to increase with time \citep[e.g.,][]{Bell_deJong_2001}; for instance, the $M_\star/L_\star$ ratio in the $B$-band can be approximated as $\propto t$ at $t\sim 1$ Gyr when a constant star formation rate (or decaying with a delay time)\footnote{
One of the most extreme SFH is the case where a galaxy forms in one burst episode at $z\rightarrow \infty$. The $M_\star/L_\star$ ratio continuously decreases with time due to death of massive stars that  are not a dominant population in mass for a standard initial mass function (IMF) \citep[e.g.,][]{Kroupa_2001,chabrier03}. However, the SFH is not applicable to our targets with active star formation inferred from strong emission lines.} is assumed \citep{Into_Portinari_2013}.
Therefore, for the purpose of deriving an upper bound of the stellar mass, we adopt a characteristic time of $t=1$ Gyr.
We use the population synthesis code {\tt STARBURST99} version v7.0.1 \citep{Leitherer_1999} to generate stellar SEDs of galaxies.
Here, we assume the Kroupa IMF (\citealt{Kroupa_2001}; $0.1-100~\msun$), the Padova isochrone models, 
and constant star formation with a duration of 1 Gyr. 
We consider two values of stellar metallicity ($Z=\zsun$ and $0.2~\zsun$) to show the metallicity dependence,
while we note that the solar-metallicity case gives a higher upper bound of the stellar mass.
We take into account dust attenuation by the extinction law of starburst galaxies \citep{calzetti00}.
The color excess of the stellar continuum is fixed to $E_{\rm s}(B-V)=0.09$,
which is calculated from the Balmer decrement of the narrow emission lines we detect in the NIRSpec spectra 
(see Section~\ref{sec: line_properties}).

This model, when scaled to the flux density in the F356W filter, results in a host mass of $M_\star = 6.0\times 10^9~\msun$.   This galaxy SED model is shown in Figure~\ref{fig:SED}a as the red curve.
Therefore, we argue that the stellar mass of the host galaxy is limited to $M_\star < 6.0\times 10^9~\msun$ for CEERS~1670
so that the stellar continuum flux density does not exceed the observed continuum level. We note that the upper bound depends significantly on the low-mass end ($m_{\star, \rm min}$) of the stellar IMF; for instance, the upper bound is reduced by a factor of $\sim 3$ for $m_{\star, \rm min}=1.0~\msun$.

\subsection{The Obscured Nature of CEERS 3210}\label{sec:SED3210}

Figure~\ref{fig:SED}b shows the prism spectrum and NIRCam photometric flux densities of CEERS~3210.
The red spectral shape with an index of $\alpha_\lambda \simeq 2.0$
at longer wavelengths can be explained either by a heavily obscured quasar (cyan) or a dusty starburst galaxy (red).
Both models require the existence of obscuring material along the line of sight:
a typical visual extinction of $A_V \simeq 3.65$ and $4.0$ for the obscured quasar and 
dusty galaxy model, respectively. 
We note that this dusty-galaxy SED is calculated with the same galaxy model as discussed in Section~\ref{sec:SED1670},
but assuming a stellar mass of $M_\star = 6\times 10^{10}~\msun$ and a different level of extinction.
However, neither of the SED models explains the excess of the observed spectrum
at $\lambda_{\rm obs}\lesssim 2~\mu$m, requiring additional blue components.

One possible explanation for the blue component is dust (and electron) scattering, 
which preserves the spectral shape of the intrinsic broad-line AGN component
\citep[e.g.,][]{Zakamska_2005}.
In fact, obscured quasars at low redshifts ($z<2.5$) tend to show optical polarization levels higher than 
those of unobscured populations \citep{Alexandroff_2018}.
The fraction of the scattered flux relative to the primary component depends on the covering factor of the scattering medium and our viewing angle. For instance, assuming that 0.6\% of the radiation flux of the intrinsic spectrum is scattered 
to our line of sight \citep[see the Torus model in][]{Polletta_2006}, the total SED is consistent with the photometric flux densities. 
Alternatively, the spectral shape of CEERS~3210 could be explained by the combination of quasar emission at short wavelengths and light from a heavily obscured starburst galaxy dominating at long wavelengths. This combination of AGN+galaxy light is shown as the blue curve in Figure~\ref{fig:SED}b.
If this is the case, CEERS~3210 would be caught in a transition stage, moving from a dust-obscured starburst to an unobscured luminous quasar by expelling gas and dust.
This model hypothesis is consistent with the dust-reddened AGN at $z=7.19$ reported by \citet{fujimoto22}, the BH mass of which is similar to that of CEERS~3210.  This would make CEERS~3210 a dusty progenitor of the luminous, unobscured quasars observed by ground-based quasar surveys.  

We can place a constraint on the host galaxy mass of CEERS~3210 following the same arguments used for CEERS~1670.  Assuming the light at longer wavelengths is entirely dominated by stellar emission and using a dust-obscured ($A_V=4.0$) version of the stellar population described Section~\ref{sec:SED1670}, we obtain an upper limit on the host mass of $M_\star \lesssim 6.0\times10^{10}~\msun$.  
It is worth noting that the unobscured galaxy SED is modeled so that it has the highest $M_\star/L_\star$ ratio,
and thus our estimate gives a conservative upper bound.
Using the hybrid quasar + dusty galaxy model does not appreciably change this upper limit as the steep spectral slope at $\lambda_{\rm obs}> 3~\mu$m is dominated by the galaxy component in the second scenario.

Nevertheless, it is difficult to distinguish these two scenarios using the current data.
Thus, multi-wavelength follow-up observations such as rest-frame infrared and far-infrared imaging are needed to further constrain the nature of CEERS~3210.
We leave a more detailed SED analysis of this source to future work.

\begin{figure}
\centering
\includegraphics[width=\linewidth]{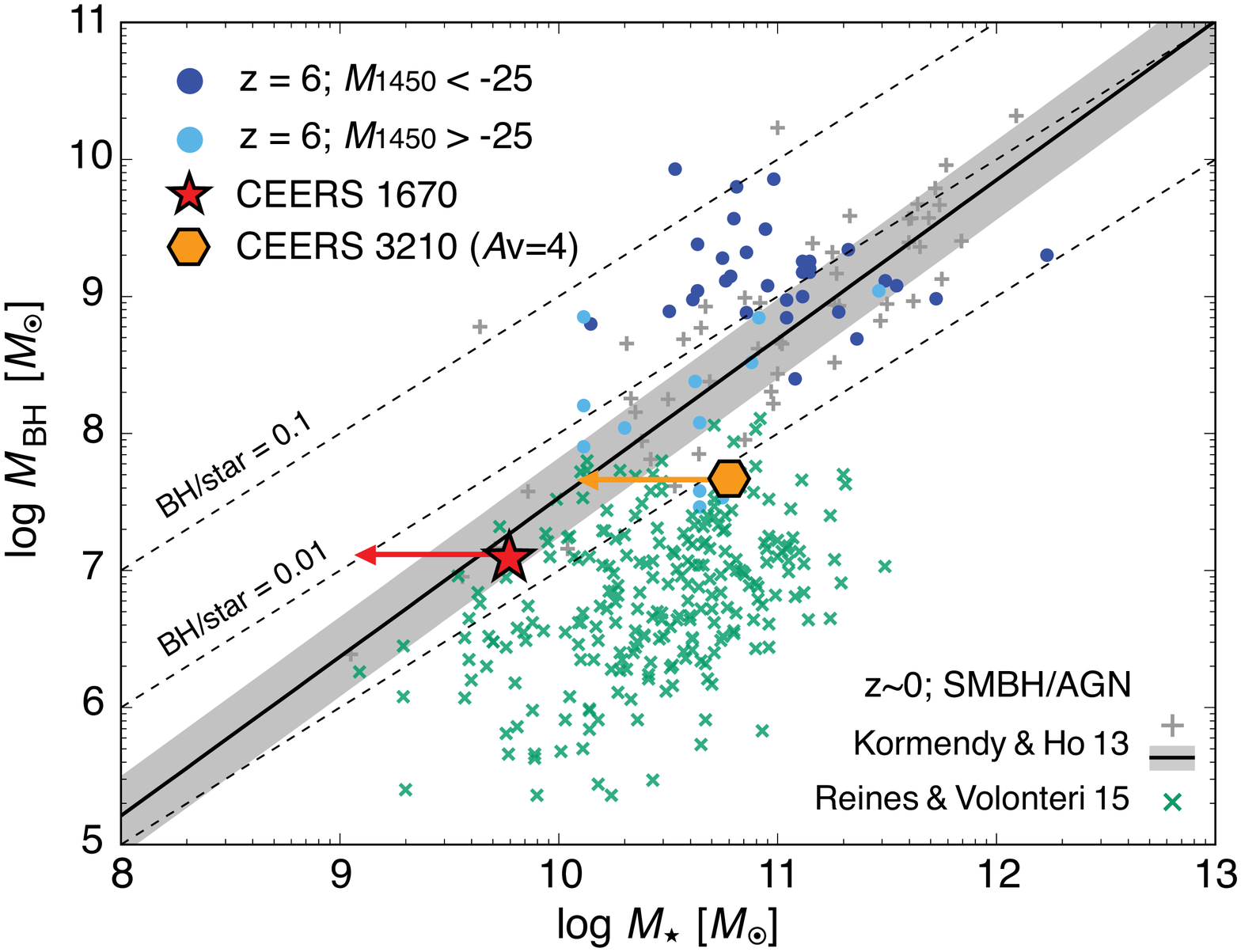}
\caption{The BH mass versus stellar mass relation of CEERS~1670 (red) and CEERS~3210 (orange; $A_V=4$).
Circle symbols show the $z >6$ quasar samples compiled by \cite{Izumi_2021}: 
brighter ones with $M_{1450} < -25$ mag (blue) and fainter ones with $M_{1450} > -25$ mag (cyan). 
The gray and green cross symbols are the observational samples in the local universe provided by 
\citet{Kormendy_Ho_2013} and \citet{Reines_Volonteri_2015}, respectively. 
The diagonal dashed lines represent $M_{\rm BH}/M_\star= 0.1$, $0.01$, and $10^{-3}$.
}
\label{fig:MM}
\end{figure}

\subsection{BH-Galaxy Coevolution at $z\simeq 5$}

The empirical relation between the masses of SMBHs and their host galaxies is considered to 
be one of the most important outcomes of their mutual evolution over the cosmic timescale 
\citep[e.g.,][]{Kormendy_Ho_2013}. 
To constrain how and when the BH-galaxy correlations were established, the $M_\star - M_{\rm BH}$ distribution
at the earliest epoch of the universe needs to be unveiled. 
The apparent location of high-$z$ quasars and their hosts also gives us more information 
on the BH growth mechanisms and their seeding processes \citep{Inayoshi_2022,Hu_2022,Scoggins_2023}.

Our first source, CEERS~1670, is a broad-line AGN with a BH mass of $M_{\rm BH}\simeq 1.3\times 10^7~\msun$
hosted in a star-forming galaxy with a stellar mass limited below $M_\star < 6.0\times 10^9~\msun$. 
Our second source, CEERS~3210, is inferred to be a heavily obscured broad-line AGN with a BH mass of 
$M_{\rm BH}\simeq 4.7\times 10^7~\msun$ (or $9.0\times 10^6~\msun$ unless it is obscured).
The host stellar mass is possibly as high as $M_\star \lesssim 6.0\times 10^{10}~\msun$
in the case of the the hybrid quasar + dusty galaxy model.

Figure~\ref{fig:MM} shows the $M_\star - M_{\rm BH}$ distribution of $z\gtrsim 6$ quasars compiled in \citet{Izumi_2021} (circle) for which the stellar mass is assumed to be the \CII-based dynamical mass.
CEERS~1670 is located at the left-bottom corner of the plane, which is uniquely separated from the $z\gtrsim 6$ quasar population already known \citep[e.g.,][]{Wang_2013,Venemans_2016,Izumi_2021}. The mass ratio of $M_{\rm BH}/M_\star > 2.4\times 10^{-3}$ for CEERS~1670 is consistent with or higher than that expected from the empirical relation seen in massive galaxies at $z=0$ \citep[black line][]{Kormendy_Ho_2013}, 
but is overmassive compared to the BH-to-galaxy mass ratio measured for nearby broad-line AGN whose virial BH masses are estimated to be as low as that of CEERS~1670 \citep{Reines_Volonteri_2015}.
On the other hand, adopting the dust-corrected BH mass and dusty-galaxy SED model, CEERS~3210 is located well 
below the empirical relation at $z\simeq 0$.
An important caveat is that the upper bound of the stellar mass can be reduced by a factor of $\simeq 3-5$
with a different stellar population and star formation history (see discussion in Section~\ref{sec:SED1670}).
Further follow-up observations would give a better estimate of the stellar mass. The existence of such an overmassive BH, if confirmed, provides us with a unique opportunity to study the early stage of the BH-galaxy assembly.

\subsection{Update of $z\sim5$ AGN Luminosity Function} \label{sec:qlf}

\begin{figure}
\includegraphics[width= \linewidth]{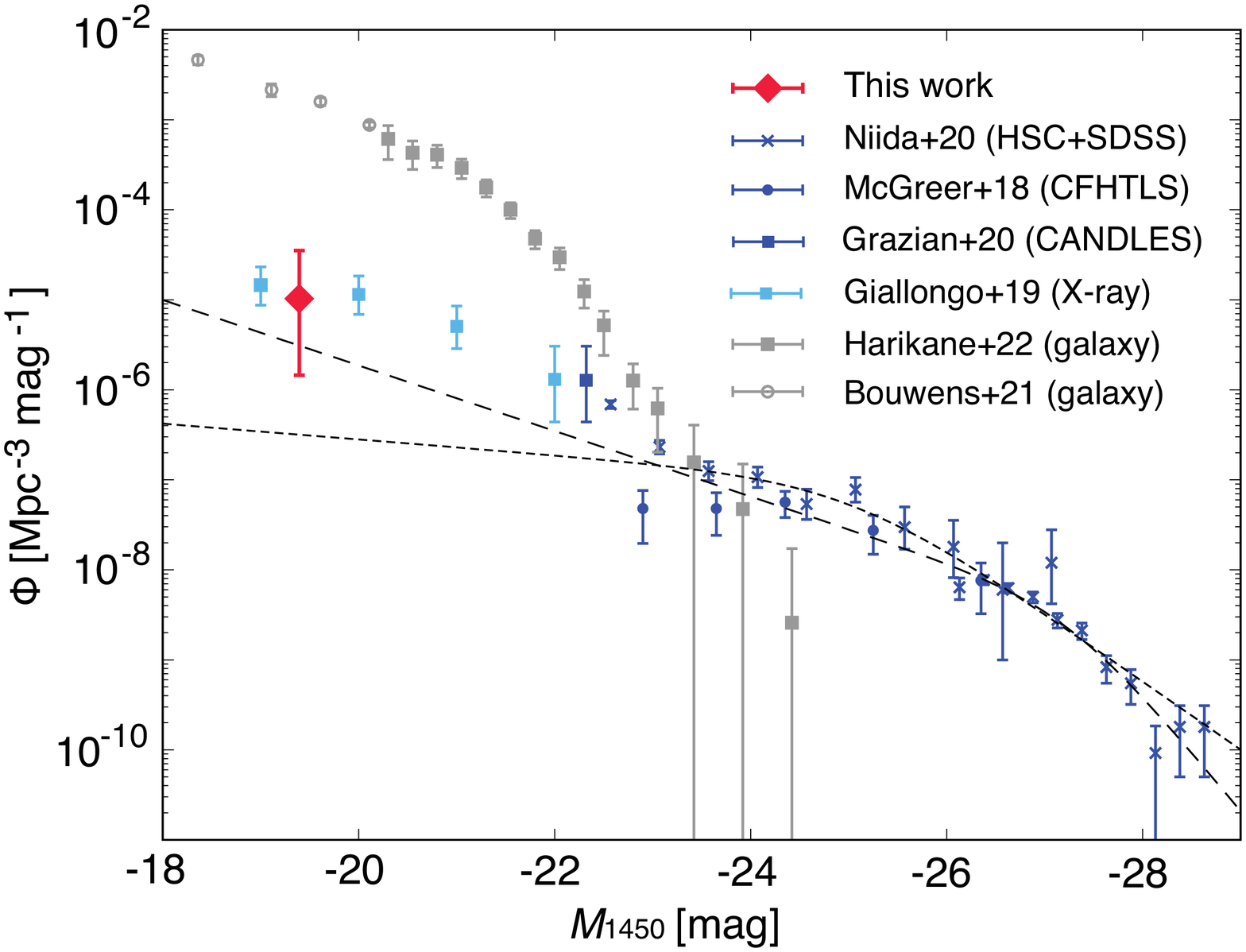}
\caption{The AGN luminosity function at $z\sim 5$ based on CEERS 1670 (red).  The 1-$\sigma$ errors have been derived using the low number count statistics by \citet{gehrels86}.
The binned luminosity function from the literature are shown for AGN \citep{McGreer_2018, giallongo19, grazian20, Niida20} and Lyman break galaxies \citep{harikane22, Bouwens21}. The short-dashed line represents the parametric luminosity function of \citet{Niida20} and the long-dashed line is from \citet{finkelstein2022b} without a correction term from a double-power law function (i.e., $\delta=0$ in their Equation~1). }
\label{fig:lfz5}
\end{figure}

We update the UV luminosity function of $z=5$ AGN from  \citetalias{onoue23}, based on the spectroscopic redshift of CEERS~1670.
We do not include CEERS~3210 in our discussion, because of its unconstrained intrinsic UV luminosity.
Following \citetalias{onoue23}, we do not aim to provide statistical constraints on the luminosity function based on our small and incomplete sample, but we rather quantify the serendipity of our discovered low-luminosity AGN at $z>5$ in the 34.5 arcmin$^2$ of the first NIRCam pointings of the CEERS survey.
Adopting the spectroscopic redshift of $z=5.24$ and the redshift interval of $\Delta z \pm0.5$, we derive the AGN number density of $\Phi = 1.07 \times 10^{-5}$ Mpc$^{-3}$  mag$^{-1}$ at the UV magnitude of $M_{1450}=-19.4$ mag.
The difference from \citetalias{onoue23} ($\Phi = 1.03 \times 10^{-5}$ Mpc$^{-3}$  mag$^{-1}$) is tiny, because the central redshift of $z=5.24$ only slightly changes from their work ($z=5.15$).
The updated luminosity function is presented in Figure~\ref{fig:lfz5}.

The faint end of the $z>5$ AGN/quasar luminosity function is a matter of debate, because low-luminosity AGN produce more ionizing photons in a certain cosmic volume than do much rarer luminous AGN, and thus its steepness is critical to infer the relative role in the cosmic reionization with respect to star-forming galaxies \citep[e.g.,][]{giallongo15, onoue17, McGreer18, Matsuoka18, giallongo19, finkelstein19, grazian20, grazian22, Niida20, kim20, kimim21, jiang22, FB22, Yung21}.
The space density that we infer suggests that low-luminosity AGN such as CEERS~1670 may in fact be common, in agreement with previous studies that have identified candidate faint quasars in relatively small survey areas \citep[e.g.,][]{fujimoto22}.
However, a complete survey of low-luminosity AGN with a well-defined selection function as well as a careful analysis of host galaxy contribution to the UV magnitudes \citep{bowler21, adams22, harikane22} is required to statistically argue the AGN abundance at the faint end, and subsequently the relative contribution of AGN to the cosmic hydrogen/helium reionization.

\section{Conclusions}

We make use of JWST NIRSpec spectroscopy from the CEERS Survey to identify two low-luminosity AGN at $z>5$ with broad H$\alpha$ emission in their spectra.  
The first source, CEERS 1670 at $z=5.242$, has a UV magnitude of $M_{1450}=-19.4\pm 0.05$, making it 2--3 dex fainter than known quasars at similar redshifts.  The source was previously identified as a candidate low-luminosity AGN based on broad-band photometry by \citetalias{onoue23}.  We measure a FWHM of $2038\pm286$ km s$^{-1}$ for the broad H$\alpha$ component, resulting in a BH mass of $M_{\rm BH}=1.3\pm0.4\times 10^{7}~\msun$, making this the least-massive BH known in the universe at the end of cosmic reionization.  

The second source, CEERS 3210 at $z=5.624$, has a blue continuum spectrum at short wavelengths ($\lambda_{\rm obs} < 3 \mu$m) and a steep spectral slope at longer wavelengths. The SED shape suggests that this source is a broad-line AGN possibly caught in a transition phase between a dust-obscured starburst and an unobscured quasar.
We measure a FWHM of $1807\pm207$ km s$^{-1}$ for the broad H$\alpha$ component, resulting in a BH mass in the range of $M_{\rm BH}\simeq 0.9-4.7 \times 10^{7}~M_{\odot}$, depending on the level of dust obscuration assumed.  

We derive upper limits on the host mass of each AGN and place constraints on the $M_\star$-$M_{\rm BH}$ relationship in the lowest mass range yet probed in the early universe.  We find the host of CEERS~1670 is limited to $M_\star < 6.0\times 10^9~\msun$, while the host mass of CEERS~3210 can be an order of magnitude larger ($6.0\times 10^{10}~\msun$) if we assume a visual extinction of $A_{V}=4.0$, as inferred from our SED fitting. The $M_{\rm BH}/M_\star$ ratio for CEERS 1670, in particular, is consistent with or higher than the empirical relationship seen in massive galaxies at $z=0$, but is overmassive compared to the BH-to-galaxy mass ratio measured for nearby broad-line AGN whose virial BH masses are estimated to be as low as that of CEERS~1670.

We examine the narrow emission-line ratios of both sources and find that their location on the BPT and OHNO diagrams is consistent with model predictions for moderately low-metallicity AGN with $Z/Z_\odot \simeq 0.2-0.4$.  The fact that neither source is X-ray detected and their emission line ratios in the BPT diagram are virtually indistinguishable from star-forming galaxies observed at similar redshifts means that their broad-line emission may be one of the few ways to detect these AGN.
Other possible approaches include diagnostics with high-ionziation lines (e.g., He and Ne) \citep{Feltre16, Nakajima_2022}.
Preselection with photometric colors may also be useful to select fast-growing BHs with $M_{\rm BH} \sim 10^{6-7}~\msun$ 
in metal-poor environments \citep{IO_2022}. 

The spectroscopic discovery of two low-luminosity, low-mass AGN at $z>5$ demonstrates the capabilities of JWST to push the BH mass limit closer to the range predicted for the BH seed population. Future work to uncover these low-luminosity AGN, which are the dominant BH population at high redshift, will be the key to further constraining their abundance and the early growth history of SMBHs and their host galaxies.

\section{Acknowledgments}

This work is supported by NASA grants JWST-ERS-01345 and JWST-AR-02446 and based on observations made with the NASA/ESA/CSA James Webb Space Telescope. The data were obtained from the Mikulski Archive for Space Telescopes at the Space Telescope Science Institute, which is operated by the Association of Universities for Research in Astronomy, Inc., under NASA contract NAS 5-03127 for JWST.  This work also made use of the Rainbow Cosmological Surveys Database, which is operated by the Centro de Astrobiología (CAB/INTA), partnered with the University of California Observatories at Santa Cruz (UCO/Lick,UCSC).

We also acknowledge support from the National Natural Science Foundation of China (12073003, 12150410307, 12003003, 11721303, 11991052, 11950410493), and the China Manned Space Project Nos. CMS-CSST-2021- A04 and CMS-CSST-2021-A06.  PGP-G acknowledges support  from  Spanish  Ministerio  de  Ciencia e Innovaci\'on MCIN/AEI/10.13039/501100011033 through grant PGC2018-093499-B-I00.

AG acknowledges financial contribution from the grant PRIN
INAF 2019 (RIC) 1.05.01.85.09: ``New light on the Intergalactic
Medium (NewIGM)'' and support from PRIN MIUR project ``Black Hole winds and the Baryon Life Cycle of Galaxies: the stone-guest at the galaxy evolution supper'', contract 2017-PH3WAT.

\end{document}